\documentclass[aps,prb,superscriptaddress,floatfix,longbibliography]{revtex4-2}

\usepackage{amsmath,amssymb} % math symbols
\usepackage{bm} % bold math font
\usepackage{graphicx} % for figures
\usepackage{comment} % allows block comments
\usepackage{textcomp} % This package is just to give the text quote '

\usepackage{enumitem}
\setlist{noitemsep,leftmargin=*,topsep=0pt,parsep=0pt}

\usepackage{xcolor} % \textcolor{red}{text} will be red for notes
\definecolor{lightgray}{gray}{0.6}
\definecolor{medgray}{gray}{0.4}

\usepackage{hyperref}
\hypersetup{
colorlinks=true,
urlcolor= blue,
citecolor=blue,
linkcolor= blue,
}

\newif\ifptitle
\newif\ifpnumber
\newcounter{para}

\ptitletrue  % comment this line to hide paragraph titles
\pnumbertrue  % comment this line to hide paragraph numbers

\newcommand{\mytitle}{Charmonium tetraquarks and pentaquarks or an additional quark?}

\begin{document}

\title{\mytitle}

\author{Scott Chapman}
%\email[]{schapman@chapman.edu}
\affiliation{Institute for Quantum Studies, Chapman University, Orange, CA  92866, USA}

\date{\today}

\begin{abstract}
Most of the exotic hadrons discovered over the last 20 years fit into the quark model as normal mesons and baryons if the existence of a seventh flavor of quark is hypothesized.  If the quark has a mass of $\sim$2.9 GeV, a charge of $-\tfrac{1}{3}$, and light-scalar-mediated decays to other quarks and $c\bar{c}$, then it is able to reproduce the mass, spin, parity, production and decay modes of most observed exotic hadrons.  Predictions are made for additional hadrons as well as new production and decay modes for observed exotic hadrons. 
\end{abstract}

\maketitle

The existence of a fourth down-type quark (charge of $-1/3$) with a mass of $\sim$2.9 GeV was recently hypothesized.  The theory involving the additional quark was presented in \cite{alternative,twisted}, where it was shown to be free of anomalies and able to reproduce experimental data that would seemingly rule out a quark with mass in that range \cite{twisted}.  The quark's production and decay mechanisms were mapped in detail to exotic hadron observations in \cite{new-quark}. 

This paper presents a listing of many of the expected mesons and baryons involving the hypothesized quark.  Each meson and baryon expected by the Quark Model (QM) is mapped to (a) an observed exotic hadron, (b) an exotic hadron seen with $<5\sigma$ evidence, or (c) a predicted hadron.  For the above 3 cases, the ``$\sigma$'' column in each table is (a) $5+$, (b) the $\sigma$ of the evidence, or (c) blank. The hypothesized additional quark is denoted by the symbol $f$.  The Name column includes both the name and the approximate mass of the hadron, and the $\Gamma$ column includes its observed or predicted width; both are in MeV.

Observed production and decay modes for observed hadrons are described in the references of the ``Ref.'' column.  The Predicted Production and Decay columns predict as-yet-unobserved processes.  For each observed hadron, three tests of these predictions can be performed: (i) observed production + predicted decay, (ii) predicted production + observed decay, and (iii) predicted production + predicted decay.  For the unobserved hadrons, only the third of these is available for testing. The proposed $f$-quark mesons and baryons are shown in the following tables.

\begin{equation}\label{fd1}
\begin{aligned}
&\textrm{Proposed }d\bar{f}\textrm{ and }f\bar{d}\textrm{ mesons}  \\
\renewcommand{\arraystretch}{1.5} 
&\begin{array}{|c|c|c|c|c|c|c|c|} \hline
\rm{QM} & \rm{Name} & \Gamma & \sigma & \rm{Ref.} &  \textrm{Predicted Production} & \textrm{Predicted Decay} & \textrm{Predicted Decay BR}\\ 
\hline
1^1 S_0 & X^0(3250) & <35 & 5+ & [4],[5] &  B^0 \to \phi X^0 & X^0 \to p\bar{p}K_S^0 & >5\% [B^0 \to \phi p\bar{p}K_S^0 ]\\
 &  &  &  &  &  B^+ \to \pi^+ X^0 & X^0 \to \pi^+\pi^- & >1\% [B^+ \to \pi^+\pi^+\pi^- ]\\
1^3 S_1 & X^0(3350) & 70 & 5+  & [6],[5] &  e^+e^- \to X^0 & X^0\to \gamma X^0(3250) &  >5 \times[X^0 \to \Lambda_c^+\bar{p}] \\
1^3 P_0 & X^0(3820)   & 30 & 2.8 & [6] & \Upsilon(1S) \to X^0(3350) X^0  & X^0 \to \Lambda_c^+\bar{p} & >5\%[\Upsilon(1S) \to 2\Lambda_c^+2\bar{p}]\\
1^3 P_1 & \chi_{c1}(3872)  & 1.2 & 5+ & [5] & B^0 \to \phi \chi_{c1} & \chi_{c1} \to K_S^0 J/\psi  & >1\% [B^0 \to \phi K_S^0 J/\psi ]\\
1^1 P_1 & T_{c\bar{c}1}^0(3900)  & 28 & 5+  & [5] &   & T_{c\bar{c}1}^0\to\pi^{\mp,0}X^{\pm,0}(3350) & > [T_{c\bar{c}1}^0\to\pi^0 J/\psi]\\
2^1 S_0 & Z_{cs}^0(3985)  & 8 & 4.6  & [7] &   &    Z_{cs}^0\to \pi^{\mp,0} X^{\pm,0}(3350) & \\
1^3 P_2 & \chi_{c2}(4014)  & 4 & 2.8  & [8] &   & \chi_{c2} \to \gamma X^0(3350) & >[\chi_{c2} \to \gamma J/\psi]\\
2^3 S_1 & T_{c\bar{c}}^0(4020)  & 13 & 5+  & [5] & B^+\to K^+ T_{c\bar{c}}^0  & T_{c\bar{c}}^0\to \mu^+\mu^- & >1\%[B^+\to K^+\mu^+\mu^-]\\
1^1 D_2 & X(4160) & 136 & 4.8 &  [5]  & B^+\to K^+ X  & X\to \mu^+\mu^- & >1\%[B^+\to K^+\mu^+\mu^-]\\
2^1 P_1 & T_{c\bar{c}}^0(4200)  & 200 &  &  & B^-\to K^- T_{c\bar{c}}^0   & T_{c\bar{c}}^0 \to \pi^{\mp,0} T_{c\bar{c}}^{\pm,0}(4020) & \\
2^3 P_1 & \chi_{c1}(4274)  & 51 & 5+  & [5] & \psi(4390)\to \gamma \chi_{c1}    &  & \\
1^3 D_1 & \psi(4320)   & 152 & 4.0 & [9] &  & \psi \to \pi^{\mp,0}T_{c\bar{c}1}^{\pm,0}(3900) & >20\%[\psi\to \pi^+\pi^- J/\psi]\\
2^3 P_2 & X^0(4350)   & 13 & 3.2 & [10] &  \psi(4390)\to \gamma X^0  &  & \\
3^3 S_1 & \psi(4390)  & 9 & 4.0 & [9] & e^+e^- \to \psi  & \psi\to \pi^\pm K^\mp J/\psi & > [\psi\to \pi^+\pi^- J/\psi]\\
3^3 P_1 & \chi_{c1}(4685)   & 126 & 5+  & [5] &   &   \chi_{c1}\to\phi X^0(3350) & >[\chi_{c1}\to\phi J/\psi]\\
3^3 P_0 & \chi_{c0}(4700)  & 87 & 5+  & [5] &   &   \chi_{c1}\to\phi X^0(3350) & >[\chi_{c1}\to\phi J/\psi]\\
\hline
\end{array}\,.
\end{aligned}
\end{equation} 
For $\chi_{c1}(3872)$ to have quantum numbers $J^{PC}=1^{++}$, it must be in the $C$ eigenstate $(f\bar{d})_+ =\tfrac{1}{\sqrt{2}}(f\bar{d}+d\bar{f})$.  Fig. 3 of \cite{Babar32502009} may contain a hint of the second predicted decay of $X^0(3250)$.  Fig. 2 of \cite{4000} may contain a hint of the predicted decay of $\chi_{c1}(3872)$, especially if a width closer to that of the $Z_{cs}^0(3985)$ is assumed for the 4$\sigma$ resonance seen at 3991 MeV.  The predicted $X(4160)$ decay overlaps with the observed $\psi(4160)$ decay \cite{psi4160}.  Together with the predicted decay of $\psi(4230)$ in the $f\bar{s}$ table below, these could provide an explanation for why the experimental $\mu^+\mu^-$ signal is approximately double the calculated amount expected for the $\psi(4160)$ decay. 

\begin{equation}\label{fu}
\begin{aligned}
&\textrm{Proposed }u\bar{f}\textrm{ and }f\bar{u}\textrm{ Mesons} \\
\renewcommand{\arraystretch}{1.5} 
&\begin{array}{|c|c|c|c|c|c|c|c|} \hline
\rm{QM} & \rm{Name} & \Gamma & \sigma & \rm{Ref.} &  \textrm{Predicted Production} & \textrm{Predicted Decay} & \textrm{Predicted Decay BR}\\ 
\hline
1^1 S_0 & X^\pm(3250) & <35 & 5+ & [4],[5] &  B^+\to K^+K^+ X^-  & X^- \to \bar{p}\Lambda & >5\% [B^+\to K^+K^+\bar{p}\Lambda]\\
 &  &  &  &  &  \Upsilon(1S)\to X^+(3350)X^-  & X^- \to \bar{p}\Lambda & >5\% [\Upsilon(1S)\to p\bar{p}\Lambda\bar{\Lambda}]\\
1^3 S_1 & X^\pm(3350) & 70 &   &  &  B^+ \to \phi X^+ & X^+\to p\bar{\Lambda}  & > 2\%[B^+ \to \phi p\bar{\Lambda}]  \\
1^3 P_0 & X^\pm(3820)   & 30 &  & &   B^+\to \pi^+\pi^- X^+  & X^+ \to p\bar{\Lambda} & >5\% [B^+\to \pi^+\pi^-p\bar{\Lambda}]\\
 &   &  &  &  & \Upsilon(4S)\to B^-(\phi X^+)   & X^+ \to K^+ J/\psi & >10\% [\Upsilon\to B^-\phi K^+ J/\psi ]\\
1^3 P_1 & T_{cc}^+(3875)  & 0.4 & 5+ & [5] & B^+\to \pi^+\pi^- T_{cc}^+  & T_{cc}^+ \to p\bar{\Lambda} & >5\% [B^+\to \pi^+\pi^-p\bar{\Lambda}]\\
 &   &  &  &  & \Upsilon(4S)\to B^-(\phi T_{cc}^+)   & T_{cc}^+ \to K^+ J/\psi & >10\% [\Upsilon\to B^-\phi K^+ J/\psi ]\\
 &  &  &  &  &  \Upsilon(1S)\to X^-(3350)T_{cc}^+  & T_{cc}^+ \to p\bar{\Lambda} & >5\% [\Upsilon(1S)\to p\bar{p}\Lambda\bar{\Lambda}]\\
 &   &  &  &  & \psi(4660)\to K^{-}T_{cc}^+   & T_{cc}^+ \to K^{+} J/\psi & >5\% [e^+e^-\to K^{-}K^+J/\psi ]\\
1^1 P_1 & T_{c\bar{c}1}^\pm(3900)  & 28 & 5+  & [5] &   & T_{c\bar{c}1}^\pm\to\pi^{\pm,0}X^{0,\pm}(3350) & >[T_{c\bar{c}1}^\pm\to\pi^\pm J/\psi]\\
2^1 S_0 & Z_{cs}^\pm(3985)  & 13 & 5.3  & [14] &   &    Z_{cs}^\pm\to \pi^{\pm,0} X^{0,\pm}(3350) & \\
1^3 P_2 & \chi_{c2}^\pm(4014)  & 4 &  &  &  \Upsilon(1S)\to X^-(3350) X^+  & X^+ \to \pi^+\pi^0 X^0(3350) & >5\% [\Upsilon(1S)\to \pi^+\pi^0 \Lambda\Lambda_c^+ 2\bar{p}]\\
2^3 S_1 & T_{c\bar{c}}^\pm(4020)  & 13 & 5+  & [5] &   & T_{c\bar{c}}^\pm\to \pi^{\pm,0}X^{0,\pm}(3350) & \sim [T_{c\bar{c}}^\pm(4055)\to \pi^\pm \psi(2S)]\\
2^1 P_1 & T_{c\bar{c}}^\pm(4200)  & 370 & 5+ & [5] &  & T_{c\bar{c}}^\pm \to \pi^{\pm,0} T_{c\bar{c}}^{0,\pm}(4020) & \\
2^3 P_0 & T_{c\bar{c}}^\pm(4250)  & 177 & 5+ & [5] &   &  & \\
2^3 P_1 & T_{c\bar{c}}^\pm(4240)  & 50 & 5+ & [5] & \bar{B}^0 \to K^- T_{c\bar{c}}^+ & T_{c\bar{c}}^+ \to K^{*+} J/\psi  &  >5\% [\bar{B}^0 \to K^-K^{*+} J/\psi]\\
1^3 D_1 & \psi^\pm(4320)   & 152 &  &  & \psi(4660)\to K^-\psi^+ & \psi^+ \to \pi^{+,0}T_{c\bar{c}1}^{0,+}(3900) & >2\%[e^+e^-\to K^-\pi^+\pi^0 J/\psi]\\
2^3 P_2 & X^\pm(4350)   & 13 & &  &  B^0\to K^{*+}X^-  & X^\pm\to \gamma X^\pm(4020) & >[X^0\to \phi J/\psi]\\
3^1 P_1 & T_{c\bar{c}1}^\pm(4430)  & 180 & 5+ & [5] &   &  & \\
3^3 P_1 & \chi_{c1}^\pm(4685)   & 126 &   &  & B^0\to K^+ \chi_{c1}^-  & \chi_{c1}^-\to\phi X^-(3350) & > [\chi_{c1}^0\to \phi J/\psi]\\
3^3 P_0 & \chi_{c0}^\pm(4700)  & 87 &   &  & B^0\to \pi^- \chi_{c0}^+ & \chi_{c0}^+\to K^+ \eta_c(1S) & >2\%[ B^0\to \pi^-K^+ \eta_c(1S)]\\
\hline
\end{array}\,
\end{aligned}
\end{equation} 
$T_{c\bar{c}}^\pm(4055)$ is assumed in this model to be the same as $T_{c\bar{c}}^\pm(4020)$.  The mass, width and $J^P$ of $T_{c\bar{c}}^+(4240)$ are consistent with those of $\chi_{c1}(4274)$ to within 2$\sigma$. Fig. 2 of \cite{Belle20183250p} may contain a hint of the first predicted decay of $X^-(3250)$.  Fig. 1 of \cite{Belle20093875p} may contain evidence of the first predicted decays of $\chi_{c0}^+(3820)$ and $T_{cc}^+(3875)$.  Fig. 6c of \cite{Babarphi3872} may contain hints of the second predicted decays of $\chi_{c0}^+(3820)$ and $T_{cc}^+(3875)$.  Fig. 7a in the supplemental material of \cite{4710} may contain a hint of the last predicted decay of $T_{cc}^+(3875)$ as well as the related decay $T_{cc}^+\to K^{*+}(700) J/\psi$.

\begin{equation}\label{fs}
\begin{aligned}
&\textrm{Proposed }s\bar{f}\textrm{ and }f\bar{s}\textrm{ Mesons}  \\
\renewcommand{\arraystretch}{1.5} 
&\begin{array}{|c|c|c|c|c|c|c|c|} \hline
\rm{QM} & \rm{Name} & \Gamma & \sigma & \rm{Ref.} &  \textrm{Predicted Production} & \textrm{Predicted Decay} & \textrm{Predicted Decay BR}\\ 
\hline
1^1 S_0 & X(3550) & <30 &   &  &  e^+e^-\to \pi^0 X & X\to \pi^0 J/\psi & >5\% [T_{c\bar{c}1}^0(3900)\to\pi^0 J/\psi] \\
1^3 S_1 & R(3760) & 22 & 5+  & [19] &   &  &  \\
1^3 P_0 & X(3960)   & 43 & 5+ & [20] & B^+\to K^+ X  & X\to \mu^+\mu^- & >3\%[B^+\to K^+\mu^+\mu^-]\\
1^3 P_1 & \chi_{c1}(4140)  & 19 & 5+ & [5] &    &  & \\
2^3 S_1 & \psi(4230)   & 49 & 5+ & [5] & B^+\to K^+ \psi  & \psi\to \mu^+\mu^- & >10\%[B^+\to K^+\mu^+\mu^-]\\
1^3 D_1 & Y(4500)  & 111 & 5+ & [21] & B^+ \to K^+ Y & Y\to\phi\psi(2S) & > 50\%[Y\to K^+K^- J/\psi]\\
2^3 P_0 & \chi_{c0}(4500) & 77 & 5+ & [22]   & B^+\to K^+ X  & X \to \phi J/\psi  & >2\% [X \to K^+ J/\psi \phi]\\
2^3 P_1 & \chi_{c1}(4600) & 133 & 5+ & [23]  &   &  & \\
3^3 S_1 & \psi(4660)   & 72 & 5+  & [5] & B^+ \to K^+\psi  & \psi \to \phi J/\psi & >1\% [B^+ \to K^+\phi J/\psi] \\
3^3 P_1 & \chi_{c1}(4900) & 255 & 5+ & [23]  &   &  & \\
\hline
\end{array}\,.
\end{aligned}
\end{equation} 
Fig 2. of \cite{3900ratio} may contain a hint of the predicted decay of $X(3550)$.   Fig. 1 of \cite{psi4160} may contain a hint of the predicted decay of $X(3960)$.

\begin{equation}\label{fc}
\begin{aligned}
&\textrm{Proposed }c\bar{f}\textrm{ and }f\bar{c}\textrm{ Mesons} \\
\renewcommand{\arraystretch}{1.5} 
&\begin{array}{|c|c|c|c|c|c|c|c|} \hline
\rm{QM} & \rm{Name} & \Gamma & \sigma & \rm{Ref.} &  \textrm{Predicted Production} & \textrm{Predicted Decay} & \textrm{Predicted Decay BR}\\ 
\hline
1^1 S_0 & X^\pm(4790) & <1 &  &  &  B_c^+\to \phi X^+   & X^+ \to  K^0_L K^+ J/\psi  & >10\%[B_c^+\to \phi K^0_L K^+ J/\psi ] \\
2^3 P_0 & X^\pm(5568) & 19 & 5+? &  [25]  &  &  & \\
3^1 S_0 & X^\pm(5660) & <30 &  &  &  pp\textrm{ (prompt)}\to X^+  & X^+ \to D_s^+ J/\psi & > 1\%[pp\to D_s^+ J/\psi]\\
3^3 P_1 & X^\pm(5810) & <30 &  &  &  pp\textrm{ (prompt)}\to X^+  & X^+ \to D_s^+ J/\psi & > 1\%[pp\to D_s^+ J/\psi]\\
\hline
\end{array}\,
\end{aligned}
\end{equation}
The question mark associated with $X^\pm(5568)$ denotes the fact that only the D0 collaboration has found evidence for it; other collaborations have searched for $X^\pm(5568)$ but have not found it. Fig. 2 of \cite{LHCb_5810} together with fig. 2a of \cite{ATLAS_5810} may contain hints of the predicted decays to $X^+(5660)$ and $X^+(5810)$.

In the prediction for $X^+(4790)$, the $K^0_L K^+$ could come from $a_0^+(980)$, $a_2^+(1320)$, $a_0^+(1450)$, or $\rho^+(1450)$.  If the $K_L^0$ is not detected, the predicted production and decay may resemble the LHCb-observed processes currently categorized as $B^+\to \phi T_{c\bar{c}\bar{s}1}^+(4000)$ then $T_{c\bar{c}\bar{s}1}^+(4000)\to K^+ J/\psi$.

\begin{equation} \label{fftable} 
\begin{aligned}
&\textrm{Proposed }f\bar{f}\textrm{ Mesons} \\
\renewcommand{\arraystretch}{1.5} 
&\begin{array}{|c|c|c|c|c|c|c|c|} \hline
\rm{QM} & \rm{Name} & \Gamma & \sigma & \rm{Ref.} &  \textrm{Predicted Production} & \textrm{Predicted Decay} & \textrm{Predicted Decay BR}\\ 
\hline
1^1 S_0 & X(6550) & 124 & 5+ & [28] &  pp\textrm{ (prompt)}\to X & X \to J/\psi J/\psi & \sim 4\%[pp\to J/\psi J/\psi]\\
1^3 S_1 & Y(6600) & \sim 300 &   &  &  e^+e^-\to X & X \to X^{\pm}(3250)X^{\mp}(3250) & >20\%[e^+e^-\to p\bar{\Lambda}\bar{p}\Lambda]\\
1^3 P_J & X(6900)  & 153 & 5+  & [5] &  Y(7250)\to\gamma X & X \to X^{\pm,0}(3250)X^{\mp,0}(3250)& >20\%[X\to\textrm{ anything}]\\
2^1 S_0 & X(7200) & 95 & 4.1 & [28] &   & X \to X^{\pm,0}(3350)X^{\mp,0}(3250)& >10\%[X\to\textrm{ anything}]\\
2^3 S_1 & X(7250) & <50 & 3.5  & [29] &  e^+e^-\to Y & Y \to X^{0}(3350)X^{0}(3350)& >20\%[e^+e^-\to p\bar{\Lambda}_c\bar{p}\Lambda_c]\\
2^3 P_J & X(7500) &  &  &   &  pp\textrm{ (prompt)}\to X & X \to \psi(2S)J/\psi  & \sim 3\%[pp\to \psi(2S) J/\psi]\\
3^1 S_0 & X(7700) &  &  &  &  pp\textrm{ (prompt)}\to X & X \to J/\psi J/\psi  & >2\%[pp\to J/\psi J/\psi]\\
\hline
\end{array}\,.
\end{aligned}
\end{equation} 
Fig. 1 of \cite{ATLAS-6600} and fig. 1 of \cite{CMS-6600} may contain hints of the predicted decays of X(7500) and X(7700). $R$ data from the Mark 1 experiment may contain a hint of the $Y(6600)$ meson (see for example fig 10 in \cite{lowR2}).

\begin{equation}\label{fb}
\begin{aligned}
&\textrm{Proposed }b\bar{f}\textrm{ and }f\bar{b}\textrm{ Mesons}  \\
\renewcommand{\arraystretch}{1.5} 
&\begin{array}{|c|c|c|c|c|c|c|c|} \hline
\rm{QM} & \rm{Name} & \Gamma & \sigma & \rm{Ref.} &  \textrm{Predicted Production} & \textrm{Predicted Decay} & \textrm{Predicted Decay BR}\\ 
\hline
1^3 S_1 & Y(8050)  & <20 & &  & e^+e^-\to Y   & Y \to J/\psi \eta_c & \\
1^3 P_1 & X(8322)  & <20 & 5+? & [32],[33] & Y(9480)\to \gamma X  &  & \\
 &   &  &  & & pp\textrm{ (prompt)}\to X  & X \to J/\psi J/\psi & \sim 1\%[pp\to J/\psi J/\psi]\\
2^3 S_1 & Y(8600)  & <80 & &  & e^+e^-\to Y  &  Y\to B^{\mp,0} X^{\pm,0}(3250) & >5\%[e^+e^-\to B^+ \bar{p}\Lambda]\\
1^3 D_1 & Y(8750)  & <80 &  &  & e^+e^-\to Y  & Y\to B^{*\mp,0} X^{\pm,0}(3250) & >5\%[e^+e^-\to B^{*+} \bar{p}\Lambda]\\
3^3 S_1 & Y(8950)  & <80 & &  & e^+e^-\to Y  & Y\to B^{*\mp,0} X^{\pm,0}(3350) & >5\%[e^+e^-\to B^{*0} \bar{p}\Lambda_c]\\
5^3 S_1 & Y(9480)  & <80 &  &  & e^+e^-\to Y  & Y\to B^{*\mp,0} X^{\pm,0}(3350) & >5\%[e^+e^-\to B^{*0} \bar{p}\Lambda_c]\\
\hline
\end{array}\,.
\end{aligned}
\end{equation}
The question mark on $X(8322)$ is due to the fact that the resonance was only seen at one run at one experiment \cite{CB1983} and not at others.  It could nonetheless have been an actual observation if a resonance 16-26 MeV heavier than the $\Upsilon(1S)$ is produced in $e^+e^-$ collisions \cite{CBR_Had}.  The $Y(9480)$ listed above is proposed to be that resonance. Hints of the predicted decay of $X(8322)$ may be contained in Fig. 1 of \cite{ATLAS-6600}, Fig. 1 of \cite{CMS-6600} and Fig. 4b of \cite{6900}.  $R$ data from the MD1 experiment may hint at some of the above $J^{PC}=1^{--}$ resonances (see for example fig 10 in \cite{lowR2}).

\begin{equation} \label{fud} 
\begin{aligned}
&\textrm{Proposed isospin 0 }fdu\textrm{ baryons in relation to their }sdu\textrm{ counterparts} \\
\renewcommand{\arraystretch}{1.5} 
&\begin{array}{|c|c|c|c|c|c|c|c|c|} \hline
sdu\textrm{ Name} & J^P & fdu\textrm{ Name} & \Gamma & \sigma & \rm{Ref.} &  \textrm{Predicted Production} & \textrm{Predicted Decay} & \textrm{Predicted Decay BR}\\
\hline
\Lambda & {\tfrac{1}{2}}^+ & \Lambda_f(3795) & <1 &  &  & B^-\to \bar{p}\Lambda_f &  \Lambda_f\to \pi^-\pi^-\Sigma_c^{++} & >5\%[B^-\to \bar{p}\pi^-\pi^-\Sigma_c^{++} ]\\
&  & &  &  &  & B^-\to \bar{p}\Lambda_f &  \Lambda_f\to \Lambda_c^{+}\pi^+\pi^-\pi^- & >2\%[B^-\to \bar{p}\Lambda_c^{+}\pi^+\pi^-\pi^-]\\
 &  &  &  &  & &  &   \Lambda_f\to \Lambda_c^{+}l^-\bar{\nu} & \\
\Lambda(1405) & {\tfrac{1}{2}}^- & \Lambda_f(4085) & <20 &  &  & B^-\to \bar{p}\Lambda_f &  \Lambda_f\to \pi^-\pi^-\Sigma_c^{++} & >1\%[B^-\to \bar{p}\pi^-\pi^-\Sigma_c^{++} ]\\
\Lambda(1520) & {\tfrac{3}{2}}^- & \Lambda_f(4200) & <20 &  & &  B^-\to \bar{p}\Lambda_f &  \Lambda_f\to \pi^-\pi^-\Sigma_c^{++} & >5\%[B^-\to \bar{p}\pi^-\pi^-\Sigma_c^{++} ]\\
\Lambda(1600) & {\tfrac{1}{2}}^+ & \Lambda_f(4280) & <100 &  &  &  &  & \\
\Lambda(1670) & {\tfrac{1}{2}}^- & P_{\psi s}^{\Lambda 0}(4338) & 7 & 5+ & [35] &  &  & \\
\hline
\end{array}\,,
\end{aligned}
\end{equation}
Fig. 7 of \cite{3795} may contain evidence of the predicted decays of $\Lambda_f(3795)$, $\Lambda_f(4085)$ and $\Lambda_f(4200)$; that paper mentions ``unexplained structures'' at 3.8 and 4.2 GeV.

\begin{equation} \label{fuu} 
\begin{aligned}
&\textrm{Proposed isospin 1 }fuu\textrm{ baryons in relation to their }suu\textrm{ counterparts} \\
\renewcommand{\arraystretch}{1.5} 
&\begin{array}{|c|c|c|c|c|c|c|c|c|} \hline
suu\textrm{ Name} & J^P & fuu\textrm{ Name} & \Gamma & \sigma & \rm{Ref.} &  \textrm{Predicted Production} & \textrm{Predicted Decay} & \textrm{Predicted Decay BR}\\
\hline
\Sigma & {\tfrac{1}{2}}^+ & \Sigma_f^+(3870) & <20 &  &  & B^-\to \bar{p}\Sigma_f^0 &  \Sigma_f^0\to \pi^-\pi^-\Sigma_c^{++} & >2\%[B^-\to \bar{p}\pi^-\pi^-\Sigma_c^{++} ]\\
\Sigma(1383) & {\tfrac{3}{2}}^+ & \Sigma_f^{+}(4065) & <50 &  &  & \Sigma_f^0(4260)\to \pi^\mp \Sigma_f^{\pm} & \Sigma_f^{\pm}\to \pi^\pm\pi^-\Lambda_c^+  & > 1\%[B^-\to \pi^+\pi^-\pi^-\Lambda_c^+\bar{p} ]\\
\Sigma(1580) & {\tfrac{3}{2}}^- & \Sigma_f^{+}(4260) & <20 &  &  & B^-\to \Sigma_f^{0} \bar{p} & \Sigma_f^{0}\to \Lambda J/\psi & > 2\%[B^-\to \bar{p}\Lambda J/\psi]\\
\Sigma(1620) & {\tfrac{1}{2}}^- & P_c^+(4312) & 10 & 5+ & [5] &  &  & \\
\Sigma(1660) & {\tfrac{1}{2}}^+ & P_c^+(4380) & 200 & 5+ & [5] &  &  & \\
\Sigma(1670) & {\tfrac{3}{2}}^- & P_\psi^{N+}(4337) & 29 & 3 & [37] &  &  & \\
\Sigma(1750) & {\tfrac{1}{2}}^- & P_c^+(4440) & 21 & 5+ & [5] &  & & \\
\Sigma(1775) & {\tfrac{5}{2}}^- & P_c^+(4457) & 6 & 5+ & [5] &  &  & \\
\Sigma(1780) & {\tfrac{3}{2}}^+ & P_c^+(4462) &  &  &  & &   & \\
\hline
\end{array}\,,
\end{aligned}
\end{equation} 
For the above table, it is assumed that $P_c^+(4457)$ and $P_c^+(4462)$ are a very close double peak analog of the $\Sigma(1775)$ and $\Sigma(1780)$.  It appears that the data could accommodate splitting $P_c^+(4457)$ into two resonances.  It is also assumed that $P_{cs}^0(4459)$ from \cite{Pcs} is the $fud$ isospin partner of the $P_c^+(4457)$ resonance(s).  Fig. 7 of \cite{3795} may contain a hint of the predicted decay of $\Sigma_f^0(3870)$.  Fig. 3 of \cite{4338} may contain a hint of the predicted decay of $\Sigma_f^{0}(4260)$.

\begin{equation} \label{fus} 
\begin{aligned}
&\textrm{Proposed }fsu\textrm{ baryons} \\
\renewcommand{\arraystretch}{1.5} 
&\begin{array}{|c|c|c|c|c|c|c|c|} \hline
J^P & fsu\textrm{ Name} & \Gamma & \sigma & \rm{Ref.} &  \textrm{Predicted Production} & \textrm{Predicted Decay} & \textrm{Predicted Decay BR}\\
\hline
{\tfrac{1}{2}}^+ & \Xi_f(3950) & <1 &  &  & \bar{B}^0 \to \bar{\Lambda}\Xi_f^0  &  \Xi_f^0 \to \Lambda_c^{+}K^- & >10\%[\bar{B}^0 \to \bar{\Lambda}\Lambda_c^{+}K^- ]\\
{\tfrac{3}{2}}^- & \Xi_f(4260) & <50 &  &  & \Xi_{ff}^- \to D^-\Xi_f^0  &  \Xi_f^0 \to D^{+}\Sigma^-(1580) & >5\%[\Xi_{ff}^- \to D^-D^{+}\Sigma^-(1580) ]\\
&  & &  &  &   &  \Xi_f^0 \to D^{+}\Sigma^-(1660) & >5\%[\Xi_{ff}^- \to D^-D^{+}\Sigma^-(1660) ]\\
\hline
\end{array}\,.
\end{aligned}
\end{equation}
Fig. 2 of \cite{Babar_fus} may contain evidence of the predicted decay of $\Xi_f(3950)$.  In the predicted decays of $\Xi_f(4260)$, if the $\Sigma$'s decay to $K^- n$ and the neutrons are undetected, the $\Xi_f^0(4260)$ could look like two meson-like resonances with effective masses near 2870 and 2900 MeV decaying to $D^+K^-$. Since the decay to $\Sigma^-(1580)$ is S-wave and the decay to $\Sigma^-(1660)$ is P-wave, the 2870 could resemble a spin 0 resonance, and the 2900 a spin 1 resonance. Unlike the other processes in this paper that involve a color-singlet scalar, the predicted $\Xi_{ff}^- \to D^-\Xi_f^0(4260)$ process may involve a color-octet scalar from the theory of \cite{twisted}.

\begin{equation} \label{fcd} 
\begin{aligned}
&\textrm{Proposed }fcd\textrm{ baryon} \\
\renewcommand{\arraystretch}{1.5} 
&\begin{array}{|c|c|c|c|c|c|c|c|} \hline
J^P & fcd\textrm{ Name} & \Gamma & \sigma & \rm{Ref.} &  \textrm{Predicted Production} & \textrm{Predicted Decay} & \textrm{Predicted Decay BR}\\
\hline
{\tfrac{1}{2}}^+ & \Xi_{fc}(5000) & <20 &  &  & \Omega_b^- \to \pi^-\Xi_{fc}^0  &  \Xi_{fc}^0 \to \Xi_c^+ K^- & >1\%[\Omega_b^- \to \pi^-\Xi_c^+ K^- ]\\
\hline
\end{array}\,.
\end{aligned}
\end{equation}
Fig. 2 of \cite{Omega_b} may contain a hint of the predicted decay of $\Xi_{fc}$.

An advantage to the model that generates the predictions of this paper is that it classifies almost all of the previously discovered charmonium tetraquarks and pentaquarks within a single framework.  But the model has significant differences from the Standard Model, so it can only be accepted after surviving very thorough testing.  An advantage of the predicted production and decay processes in this paper is that most of them can be tested using data already collected in previous or ongoing experiments.

\end{document}